\documentclass[aps,prl,twocolumn,showpacs,amsmath,amssymb,groupedaddress]{revtex4}

\usepackage{graphicx} % Include figure files
\usepackage{dcolumn}% Align table columns on decimal point
\usepackage{bm}% bold math
\begin{document}
%\preprint{}

\title{Controlling coherence using the internal structure of hard $\pi$ pulses}


\author{Yanqun Dong, R. G. Ramos, Dale Li, and S. E. Barrett}
\email[e-mail: ]{sean.barrett@yale.edu}
\affiliation{Department of Physics, Yale University, New Haven, Connecticut 06511}
\homepage[web: ]{http://opnmr.physics.yale.edu}

% \affiliation command. The \affiliation command should follow the other information
%\author{}
%\thanks{}


\date{\today}

\begin{abstract} 
The tiny difference between hard $\pi$ pulses and their delta-function approximation can be exploited to control coherence.
Variants on the magic echo that work despite a large spread in resonance offsets are demonstrated using the zeroth- and first-order
average Hamiltonian terms, for $^{13}$C NMR in C$_{60}$. The $^{29}$Si NMR linewidth of Silicon has been reduced by a factor of about $
70,000$ using this approach, which also has potential applications in magnetic resonance microscopy and imaging of solids.
\end{abstract}

%\keywords{}

\maketitle
% References should be done using the  \cite, \ref, and \label commands

%%%%%%%%%%%%%%%%%%%%%%%%%%%%%%%%%%%%%%%%%%%%%%%%%%%%%%%%%
%%%%%%  B E G I N    B O D Y    T E X T  %%%%%%%%
%%%%%%%%%%%%%%%%%%%%%%%%%%%%%%%%%%%%%%%%%%%%%%%%%%%%%%%%%
%Nuclear magnetic resonance experiments rest upon a solid theoretical foundation  \cite{slichter, abragam, mehringText,
%ernstText}. 

In magnetic resonance, a control pulse is ÔhardÕ if the pulse amplitude is much greater than the spectral linewidth and
any resonance offset; hard pulses are often approximated as instantaneous delta-functions  \cite{Slichter, Mehring,
Ernst}.  The corrections to this picture are quite small for a single hard pulse, but they can lead to surprisingly
large effects  \cite{DalePRL, DalePRB} in important nuclear magnetic resonance (NMR) experiments that use many $\pi$ pulses,
such as the Carr-Purcell-Meiboom-Gill (CPMG) experiment  \cite{Purcell, Gill}. Even though CPMG uses a very simple
pattern of pulses  \cite{Slichter}, Coherent Averaging Theory  \cite{WaughPR} shows that the zeroth- and first-order
correction terms arising from non-zero pulse duration are quite complicated  \cite{DalePRL, DalePRB}, making a
quantitative prediction of their effects very difficult.  

In this Letter, building upon our earlier results  \cite{DalePRL, DalePRB}, we design more complicated pulse sequences,
and show that much simpler approximate Hamiltonians can quantitatively explain the experiments.  
This shows that the small difference between hard $\pi$ pulses and their delta-function approximation can be put to good
use, enabling new classes of spin echoes which have promising applications in NMR, magnetic resonance imaging (MRI) or microscopy of
solids, and related spectroscopies.

The NMR data in Figs. 1-4 of this paper were obtained with powder samples (C$_{60}$ or Silicon doped with Sb ($10^{17}/$cm$^{3}$))
at room temperature, in $B_{ext}$=12 Tesla. Both samples are well-approximated as a single species of spin I=1/2 nuclei
($^{13}$C or $^{29}$Si), coupled together by the like-spin dipolar interaction  \cite{DalePRL, DalePRB}. 
For a mesoscopic cluster of N-spins, the Hamiltonian in the rotating frame is
$\mathcal{H}_{int}\!\!=\!\!\mathcal{H}_Z + \mathcal{H}_{zz}$, where a net resonance offset 
$\left(\Omega_{z}^{net}\!\!=\!\Omega_{offset}^{global}\!+\!\Omega_z^{loc}\right)$ gives rise to the Zeeman term
$\mathcal{H}_Z\!\!\!=\!\!\!\Omega_{z}^{net}I_{z_T}$, 
and the secular part of the homonuclear dipolar coupling  \cite{Slichter, Mehring, Ernst} is $\mathcal{H}_{zz} \!\!\!=
\!\!\! \sum_{j>i}^{N}\! B_{ij}(3I_{z_{i}}I_{z_{j}}-\vec{I}_{i}\cdot\vec{I}_{j})$. Our macroscopic powders are similar to
an ensemble of N-spin clusters, with distinct $\Omega_{z}^{loc}$  values in different clusters due to bulk diamagnetism
 \cite{DalePRB}. The resulting Zeeman line broadening dominates the spectrum's full width at half maximum (FWHM), which
was only about
2 ppm (e.g., the $^{13}$C ($^{29}$Si) spectrum's measured FWHM = 260 Hz (200 Hz), while the calculated dipolar FWHM=38
Hz (88 Hz)). The rf pulses used were unusually hard (e.g., the pulse strength $\omega_{1}/2\pi \approx 25$ kHz (16.4
kHz) was about 100 (82) times the $^{13}$C ($^{29}$Si) linewidth, with a 128.56 MHz (101.56 MHz) Larmor
frequency  \cite{Slichter}).  Low coil filling factors  \cite{Slichter} ($<$8\% for $^{13}$C data and $\sim$40\% for
$^{29}$Si data) made the rf pulses very uniform across the samples.

%% FIGURE 1 %%%%%%%%%%%%%%%%%%%%%%%%%%%%%%%%%%%%%%%%%%
\begin{figure}
\includegraphics[width=3.0 in]{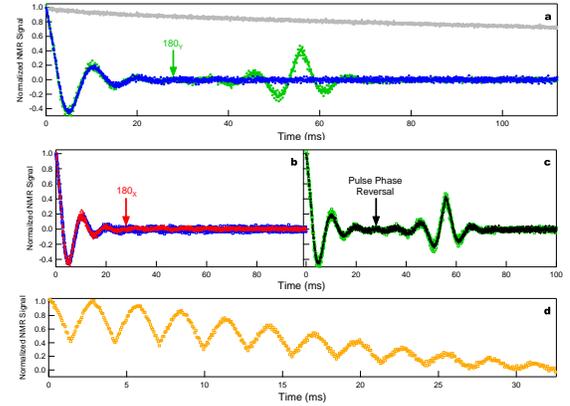}
\caption{\label{fig1}(color online) Sample C$_{60}$. (a) Comparison of CPMG (grey) to APCPMG (blue). Inserting a single
flip-$180_Y$ pulse into APCPMG induces an echo of the echo train (green). (b) Inserting a single $180_X$ (red) has no
effect (blue). (c) Reversing the APCPMG phase pattern, $90_X\!-\!{\{-Y,Y\}}^{200}\!-\!{\{Y,-Y\}}^{600}$, at the point indicated
has the same effect (black) as inserting a single $180_Y$ pulse (green). (d) A CPMG of the echo train is induced by
using $90_X\!-\!{\{-Y,Y\}}^{10}\!-\!({\{Y,-Y\}}^{20}\!-\!{\{-Y,Y\}}^{20})_{repeat}$. 
For (a-d), $\tau=25\mu s$, $\Omega_{offset}^{global}\!=0$, $\alpha\approx 0.71$,  and only the peak of
each echo is shown.  The signals in Figs. 1-4 are normalized to the amplitude of the C$_{60}$ or
Si:Sb FID signal.}
\end{figure}
%% FIGURE 1 %%%%%%%%%%%%%%%%%%%%%%%%%%%%%%%

The open grey squares in Fig. \ref{fig1}(a) show the amplitude of each peak in a long-lived train of spin echoes  \cite{Hahn}
generated by the Carr-Purcell-Meiboom-Gill (CPMG) experiment, $90_X-{\{Y,Y\}}^N$, where the first pulse is a $90^\circ$
rotation about the X-axis in the rotating frame  \cite{Purcell, Gill}.  The block $\{Y,Y\}$, repeated N times, represents
the sequence $(\tau-180_Y-2\tau-180_Y-\tau)$ where the $180^\circ$ rotations are about the Y-axis, and echoes are
acquired in the $2\tau$ time interval after every $180^\circ$ (or $\pi$) pulse  \cite{Slichter, Mehring, Ernst}. In contrast,
the train of spin echoes quickly decays to zero (Fig. \ref{fig1}(a), blue) for Alternating-Phase CPMG (APCPMG)
$90_X-{\{-Y,Y\}}^N$.

To understand this dramatic difference, we apply Coherent Averaging Theory  \cite{WaughPR} to the repeating block
$\{\phi_1,\phi_2\}$, with $180^\circ$ pulses of duration $t_p$ about the $\phi_1$ or $\phi_2$ axis, and cycle time
$t_c=4\tau+2t_p$. Short $t_c$ is used throughout this paper, so it is a good approximation to keep just the first two terms
$\bar{\mathcal H}^{(0)}+\bar{\mathcal H}^{(1)}$ in the Magnus expansion   \cite{DalePRB}. The
$\{Y,Y\}$ block has  \cite{DalePRB} $\bar{\mathcal H}^{(0)}_{\{Y,Y\}}\!\!\!=\!\!\!\alpha \mathcal H_{zz}-\beta\mathcal
H_{yy}\!\!\equiv\!\!\mathrm{H}$, while the $\{-Y,Y\}$ block has a slightly different form: $\bar{\mathcal
H}^{(0)}_{\{-Y,Y\}}\!\!=\!\!\mathrm{H}-\lambda\Omega_z^{net} I_{x_T}$, where $\alpha = \frac{4\tau}{t_c}$,
$\beta=\frac{t_p}{t_c}$, $\lambda = \frac{4t_p}{\pi t_c}$, and $\mathcal{H}_{\sigma\sigma}=\sum_{j>i}^{N}\!
B_{ij}(3I_{\sigma_{i}}I_{\sigma_{j}}-\vec{I}_{i}\cdot\vec{I}_{j})$ for $\sigma$=x, y, or z. The extra term  $-\lambda
\Omega_z^{net} I_{x_T}$ looks like a constant transverse field in the X-direction, which, when acting alone, causes
spins to nutate  \cite{Slichter} in the Y-Z plane in a manner we define as clockwise (CW). Variation in $\Omega_z^{net}$
values across the macroscopic sample leads to a spread in precession angles that causes signal decay.  In the well known
free induction decay (FID), $T_2^\star$  arises from a spread in $\Omega_z^{net}$ of the original Zeeman Hamiltonian. 
By analogy, the rapid decay of the spin echoes produced by $90_X-{\{-Y,Y\}}^N$ (Fig. \ref{fig1}(a), blue) can be thought of as an
`FID of the echo train'.

Attempting to undo this $T_2^\star$-like decay, we insert a single $180_Y$ pulse into the APCPMG sequence,
$90_X-{\{-Y,Y\}}^{N_1}-180_Y-{\{-Y,Y\}}^{N_2}$, which produces a striking `echo of the echo train' (Fig. \ref{fig1}(a),
green).  Although this looks like a conventional Hahn echo  \cite{Hahn}, the signal actually extends over more than 800
individual spin echo peaks. The dephasing caused by $-\lambda \Omega_z^{net} I_{x_T}$ (CW precession) during the
$N_1t_c$ is followed by counter-clockwise (CCW) precession caused by $+\lambda \Omega_z^{net} I_{x_T}$, and this
rephasing leads to the echo of the echo train when $N_2\!\!=\!\!N_1$. When a single flip-$180_X$ is used instead, no
echo of the echo train (Fig. \ref{fig1}(b) red) is seen, as predicted by our model, because a perfect rotation along the
X-axis does not change the sign of the $-\lambda \Omega_z^{net} I_{x_T}$ term.  On the other hand, the echo of the echo
train (Fig. \ref{fig1}(c) black) is recovered if the flip-$180_X$ is removed and the phase pattern in the second repeating
block is reversed from $\{-Y,Y\}$to $\{Y,-Y\}$, since  \cite{DalePRB} $\bar{\mathcal
H}^{(0)}_{\{Y,-Y\}}\!\!\!=\!\!\mathrm{H}\!\!+\!\!\lambda\Omega_z^{net} I_{xT}$, compared to $\bar{\mathcal
H}^{(0)}_{\{-Y,Y\}}\!\!\!=\!\!\mathrm{H}\!\!-\!\!\lambda\Omega_z^{net} I_{xT}$. In Fig. \ref{fig1}(c), the phase reversal of 1200 hard
$\pi$ pulses yields a signal indistinguishable from that induced by the single flip-$180_Y$, as predicted by our model. In
contrast to this model, taking the limit of delta-function pulses ($t_p\to 0$ )would kill  \cite{DalePRL,DalePRB} the
transverse field terms in $\bar{\mathcal H}^{(0)}\!+\!\bar{\mathcal H}^{(1)}$ exploited here and throughout the rest of
the paper. Figure \ref{fig1}(d) shows that the approach of Fig. \ref{fig1}(c) can be repeated, creating multiple echoes in the envelope of
individual spin echo peaks, or a `CPMG of the echo train'.  However, the signal does decay, since the sign of the term
$\mathrm{H}$ is never reversed in Fig. \ref{fig1}. To beat this decay, we use an approach inspired by the magic
echo   \cite{Slichter,WaughPRL,WaughPRB}.

\begin{figure}
\includegraphics[width=2.8 in]{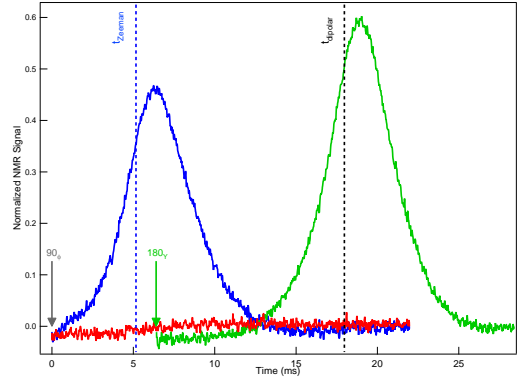}
\caption{\label{fig2}(color online). 
Sample C$_{60}$. Three experiments inspired by the magic echo  \cite{WaughPRL,WaughPRB}, which all start with
${\{-X,X\}}^N$, have distinctly different results.  With $90_{-X}$ following the repeating block, no magic echo forms
(red); with $90_{+X}$ following the repeating block, a large echo emerges (blue); when applying a $180_Y$ pulse at time
$t_{f_1}$ after the burst of the failed sequence (red), an optimized echo is achieved (green). Here, N=200, $\tau=50\mu s$, 
$\Omega_{offset}^{global}\!=0$, and $\alpha\approx0.83$.}
\end{figure}

In the original magic echo  \cite{WaughPRL,WaughPRB,McDowell}, a continuous rf field in the transverse plane picks out the
part of the dipolar coupling that is secular in the strong transverse field  \cite{Slichter}. In the $\{-X,X\}$ block \cite{DalePRB}, the
effective field $\lambda \Omega_z^{net} I_{y_T}$ in ${\mathcal H}^{(0)}_{\{-X,X\}}$ could play the same role, as first
proposed by Pines and Waugh  \cite{WaughJMR} for a single value of $\Omega_z^{net}$. Figure \ref{fig2} provides experimental
support for their prediction, even though the weakness of the effective transverse field makes it hard to justify
the second averaging analysis  \cite{WaughJMR,WaughJCP}. In addition, the spread in $\Omega_z^{net}$ across the
macroscopic sample has non-trivial consequences, as shown by the different effects (Fig. \ref{fig2}) of the two ÒburstsÓ,
$\{-X,X\}^N-90_{\pm X}$, followed by a free evolution of duration $t_{free}$. Using our model, the unitary operators are
$e^{-\frac{i}{\hbar}\left(\mathcal H_{zz}+\Omega_z^{net} I_{zT}\right)t_{free}}
e^{-\frac{i}{\hbar}\left(\frac{-\left(\alpha-\beta\right)}{2}\mathcal H_{zz} \mp \lambda \Omega_z^{net} I_{zT}\right)Nt_c}
\mathcal U_{90_{\pm X}}$, where $\pm X$ is the $90^\circ$ pulse phase and $\alpha \!\!>\!\!\beta$ for our experiments \cite{RonaPRB}.

%% FIGURE 2 %%%%%%%%%%%%%%%%%%%%%%%%%%%%%%%%%%%%%%%%%%
%% FIGURE 2 %%%%%%%%%%%%%%%%%%%%%%%%%%%%%%%%

For the -X choice, the Zeeman phase wraps in a CCW manner both during and after the burst, which spoils the magic echo
that would otherwise form during the free evolution period (Fig. \ref{fig2}, red).  For the +X choice, both Zeeman and
dipolar terms switch from CW phase wrapping in the burst to CCW phase unwrapping during the free evolution period,
resulting in a large echo (Fig. \ref{fig2}, blue).  This echo is not optimized, since the refocusing time is different
for the dipolar and Zeeman phases ($t_{dipolar}=\left(\alpha-\beta\right)Nt_c/2$, $t_{Zeeman}=\lambda Nt_c$).  An
optimized echo (Fig. \ref{fig2}, green) is generated if we apply a $180_Y$ at time
$t_{f_1}=\left(\frac{\alpha-\beta-2\lambda}{4}\right)Nt_c$ after the failed sequence $\{-X,X\}^N-90_{-X}$ (Fig. \ref{fig2},
red). This sequence aims to synchronize the refocusing times of the dipolar and Zeeman phases by using the fact that a
$180_Y$ pulse flips the sign of the Zeeman term but does not change the dipolar term. The measured echo happens at a
slightly different time, due to terms ignored in this model \cite{RonaPRB}.

Compared to the original magic echo  \cite{WaughPRL,WaughPRB}, which works best if $\Omega_z^{net}\!\!=\!\!0$, sequences
based on the $\{-X,X\}^N$ block have several clear differences: both Zeeman and dipolar phases are wrapped during the
burst, a $90_{\pm X}$ is used instead of the $90_Y$, and the $2\tau$ gaps in between the $\pi$ pulses of the $\{-X,X\}$
block simplify implementation.

%% FIGURE 3 %%%%%%%%%%%%%%%%%%%%%%%%%%%%%%%%%%%%%%%%%%
\begin{figure}
\includegraphics[width=2.7 in]{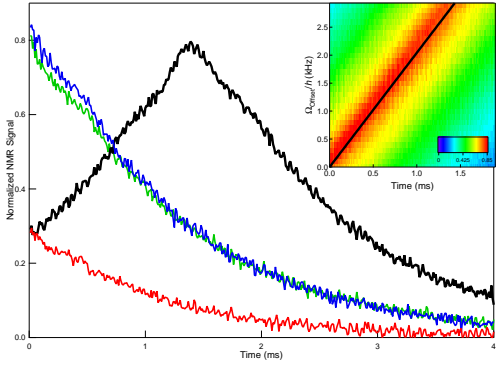}
\caption{\label{fig3}(color online). 
Sample C$_{60}$. Using $\Omega_z^{net,\pm}=\Omega_z^{loc}\pm\Omega_{offset}^{global}$, 
the quadratic echoes produced by $\{X,X\}^{\frac{N}{2}}\{-X,-X\}^{\frac{N}{2}}-90_Y-t_{free}$: green
($\nu_{offset}= 0$ Hz), black ($\nu_{offset}= -3$ kHz), differ from the linear echoes produced by
$\{-X,X\}^{\frac{N}{2}}\{X,-X\}^{\frac{N}{2}}-90_X-t_{free}$: blue ($\nu_{offset}= 0$ Hz), red
($\nu_{offset}= -1$ kHz), where $\Omega_{offset}^{global}=-h\nu_{offset}$. Only the black
echo shifts to the right.  (Inset) Image plot of 31 quadratic echoes for 0 Hz $\le\Omega_{offset}^{global}/h\le$ 3
kHz, in steps of 100 Hz. The black trend line shows our predicted Zeeman refocusing time. Here, N=100, $\tau=10\mu s$, and $\alpha\approx0.5$.}
\end{figure}
%% FIGURE 3%%%%%%%%%%%%%%%%%%%%%%%%%%%%%%%%

The $\{\phi_1,\phi_2\}$ blocks used so far have an effective transverse field term in $\bar{\mathcal
H}^{(0)}_{\{\phi_1,\phi_2\}}$. However, for the
block $\{X,X\}$, $ \bar{\mathcal H}^{(0)}_{\{X,X\}}\!\!=\!\!\alpha \mathcal H_{zz}\!\! -\!\!\beta \mathcal H_{xx}$, and so the
first transverse field term is in $\bar{\mathcal H}^{(1)}_{\{X,X\}}\!\!=\!\!-\left(\kappa\Omega_z\right)^2
I_{x_T}\!\!+\!\bar{\mathcal H}^{(1),non-I_{x_T}}_{\{X,X\}}$, where
$\kappa^{2}\!\!=\!\!t_p\left(8\tau+2t_p\right)/\left(2t_c\hbar\pi\right)$ \cite{DalePRB}.  In principle, despite its smaller size, 
the quadratic transverse field term of $\bar{\mathcal H}^{(1)}_{\{X,X\}}$ could be exploited just like the linear transverse field term found in
$\bar{\mathcal H}^{(0)}_{\{-X,X\}}$.  In practice, however, $\{X,X\}^N$ is a poor nutation experiment,
since $\bar{\mathcal H}^{(1),non-I_{x_T}}_{\{X,X\}}$ causes rapid signal decay. Inspired by the Rotary Echo
experiment   \cite{Solomon}, we tried replacing ${\{X,X\}}^N$ with the
composite block ${\{X,X\}}^{\frac{N}{2}}{\{-X,-X\}}^{\frac{N}{2}}$ because $ \bar{\mathcal
H}^{(0)}_{\{-X,-X\}}\!\!=\!\!\bar{\mathcal H}^{(0)}_{\{X,X\}}$ and $\bar{\mathcal H}^{(1)}_{\{-X,-X\}}\!\!=\!\!-
\bar{\mathcal H}^{(1)}_{\{X,X\}}$  \cite{DalePRB}, and we managed to recover most of the original signal. We thus infer \cite{RonaPRB} that the net effect of
${\{X,X\}}^{\frac{N}{2}}{\{-X,-X\}}^{\frac{N}{2}}$ is well-approximated by the much simpler unitary operator 
$ e^{\!-\frac{i}{\hbar}(-\frac{1}{2}\mathcal H_{xx} +( \kappa \Omega_z^{net,-})^2 I_{xT})\frac{Nt_c}{2}} e^{\!-\frac{i}{\hbar}(-\frac{1}{2}\mathcal H_{xx} -(\kappa \Omega_z^{net,+})^2 I_{xT})\frac{Nt_c}{2}}$, 
where we allow for different $\Omega_z^{net,\pm}$ during
${\{\pm X,\pm X\}}^{\frac{N}{2}}$.

To test our model, we use phase-coherent frequency jumping $\Omega_z^{net,\pm}\!\!=\!\!\Omega_z^{loc}\pm\Omega_{offset}^{global}$
($\Omega_{offset}^{global}\!\!\!\ge\!\!\!0$) during the burst
${\{X,X\}}^{\frac{N}{2}}{\{-X,-X\}}^{\frac{N}{2}}\!\!-\!\!90_Y$, followed by $\Omega_{offset}^{global}\!\!=\!\!0$ during free
evolution, leading to \cite{RonaPRB} the unitary operator 
$e^{\!-\frac{i}{\hbar}(\mathcal H_{zz}\!+\Omega_z^{loc} I_{zT})t_{free}} e^{\!-\frac{i}{\hbar}(-\frac{1}{2}\mathcal H_{zz}\! -( 2\kappa^2 \Omega_{offset}^{global})\Omega_z^{loc}I_{zT})Nt_c} \mathcal U_{90_Y}$. 
Increasing $\Omega_{offset}^{global}$ increases the Zeeman dephasing during the burst, pushing the quadratic echo peak out to later in $t_{free}$  (Fig. \ref{fig3},
green and black). The inset of Fig. \ref{fig3} shows the strong agreement between the Zeeman refocusing time predicted by
our model (black trend line) and the quadratic echo peak measured in our experiments over a range of
$\Omega_{offset}^{global}$. In contrast, the corresponding ÒlinearÓ sequence
${\{-X,X\}}^{\frac{N}{2}}{\{X,-X\}}^{\frac{N}{2}}-90_X$ has its largest signal just after the burst, for all
$\Omega_{offset}^{global}$ (Fig. \ref{fig3}, blue and red), as predicted in our model \cite{RonaPRB}.

Controlling both dipolar and Zeeman phase wrapping using $\bar{\mathcal H}^{(1)}_{\{\phi_1,\phi_2\}}$ is an unusual aspect
of the quadratic echo.  As one use of this, we designed a composite block with no net dipolar evolution over duration of $6\Delta$,  
$\left(\Delta\!\!+\!\!\delta\right)\!\!-\!\!90_{\psi_1}\!\!-\!\!{\{X,X\}}^{\frac{N}{2}}{\{-X,-X\}}^{\frac{N}{2}}\!\!-\!\!90_{\psi_2}\!\!-\!\!\left(\Delta\!\!-\!\!\delta\right)$, 
which we refer to as $\{N,\delta,\psi_1,\psi_2\}$, with $\Delta= Nt_c/4$, $|\delta|\le \Delta$, and
$\psi_i=\pm Y$ for i=1,2. For constant $\Omega_z^{net}$, the unitary operator is $\mathcal U_{180_Y}
e^{-\frac{i}{\hbar}\left(\Omega_z^{net}I_{z_T}\right)\left(+2\delta\right)}$ for $\psi_1=\psi_2$, and
$e^{-\frac{i}{\hbar}\left(\Omega_z^{net}I_{z_T}\right)\left(+2\Delta\right)}$ for $\psi_1\neq\psi_2$ \cite{RonaPRB}. While similar
effective operators were previously demonstrated  \cite{Matsui} using magic sandwich echoes for
$\parallel\!\!\mathcal H_Z\!\!\parallel\ll\parallel\!\!\mathcal H_{zz}\!\!\parallel$, our approach works in the complimentary regime
$\parallel\!\!\mathcal H_Z\!\!\parallel\ge\parallel\!\!\mathcal H_{zz}\!\!\parallel$, where the scales are
calculated using  \cite{Mehring,DalePRB} $\parallel \!\!A\!\!\parallel^{2}\equiv Tr(A^{\dagger}A)$.
In particular, the $\{N,\delta,\psi_1,\psi_2\}$ sequence is still effective even when there is a large spread in $\Omega_z^{net}$ values across the sample \cite{RonaPRB}.

%% FIGURE 4 %%%%%%%%%%%%%%%%%%%%%%%%%%%%%%%%%%%%%%%%%%
\begin{figure}
\includegraphics[width=3.0 in]{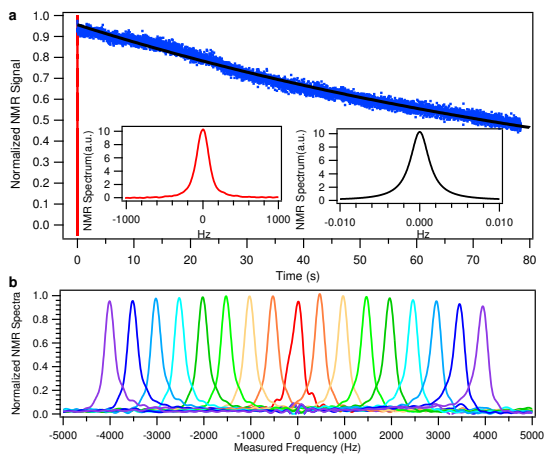}
\caption{\label{fig4}(color online). (a) Sample Si:Sb. The $^{29}$Si time-suspension data using the sequence
$90_X-\{2,0,-Y,-Y\}^{84000}$ with $\tau=60\mu s$, $\nu_{offset}$= 2.5 kHz (blue) and corresponding fitting curve (black),
extend far beyond the normal $^{29}$Si FID with $\nu_{offset}$= 0 Hz (red). (a, Inset) The 200 Hz normal spectrum (red) is narrowed to 0.003 Hz
(black, Fourier transformation of the fitting curve), centered at $\nu_{offset}$. (b) Sample C$_{60}$. Reproduction of a
top-hat lineshape using sequence $90_X-\{2,t_0,-Y,-Y\}-\{2,0,-Y,Y\}^{30}$ with $\tau=22\mu s$, and $t_0$=0.  Each trace
is the measured spectrum of a pseudo-FID with different $\nu_{offset}$ , for -4 kHz$\le\nu_{offset}\le$+4 kHz in steps
of 500 Hz, covering the range $2\pi|\nu_{offset}|/\omega_1\le 16\%$. To obtain this full bandwidth, the pseudo-FID
interleaves a second data set using the same sequence, but with $t_0=-(\frac{\Delta}{2}+\frac{1}{2\omega_1})$.}
\end{figure}
%% FIGURE 4%%%%%%%%%%%%%%%%%%%%%%%%%%%%%%%%

Our model predicts that both Zeeman and dipolar phase are refocused after each $\{N,0,\psi_1,\psi_1\}$ block, yielding a
Ôtime-suspensionÕ sequence  \cite{Matsui}.  Indeed, in Si:Sb, our sequence pushes the
decay time from $T_2^\star \approx 1.6$ ms out to  $T_2^{effective}\approx110$ seconds, or about $10^{10}$ periods of Larmor precession (Fig. \ref{fig4}(a),
blue), quite close to the spin-lattice relaxation time, $T_1=290$ seconds. The normal linewidth is thus reduced by a
factor of about 70,000 ((Fig. \ref{fig4}(a), Inset).  

Eliminating dipolar dephasing in order to measure $\Omega_z^{net}$ in applied magnetic field gradients enables the
MRI  \cite{Matsui,Garroway,Blumich} or MR microscopy  \cite{Mansfield} of solids. Measuring the
spectrum in a field gradient is the first step toward imaging using the back-projection
technique  \cite{Slichter,Blumich}. Figure \ref{fig4}(b) shows a faithful reproduction of an input top-hat spectrum, where
each spectrum is the Fourier transformation of the pseudo-FID resulting from two interlaced data sets \cite{RonaPRB}. Note that both
the signal amplitude and the $\nu_{offset}$ values have been quite accurately reconstructed in this approach. Compared
to existing approaches for the MRI of solids \cite{Matsui,Garroway,Blumich}, our approach does not need to switch off the
applied Zeeman gradient inside the bursts, which enables the application of large field gradients at moderate cost. It
should also be possible to implement standard frequency- and phase-encoding methods using this
approach \cite{Slichter,Blumich}. Since pulse strength varies across a big sample, the uniform pulse assumption of our
model is a potential concern.  Experimentally, an intentional uniform misadjustment of all pulse angles leads to similar
MRI top-hat lineshapes and to similar line-narrowing performance, suggesting that these two sequences are robust \cite{RonaPRB}.

Our sequences may help in the study of some important biomaterials, since the $\mathcal H_{int}$ assumed here is very similar to that of
$^{31}$P in bones and teeth  \cite{Ackerman,Glimcher}. Preliminary results are encouraging \cite{RonaPRB}. These sequences also have
potential applications in proton ($^1$H) NMR. While the dipolar linebreadth dominates most $^1$H spectra, a large
$\Omega_{offset}^{global}$ can be used to reach the $\parallel\mathcal H_Z\parallel\ge\parallel\mathcal H_{zz}\parallel$
limit of our model, as demonstrated in our preliminary results on Adamantane \cite{RonaPRB}.  Future work will use
microcoils  \cite{Lauterbur,Kentgens,Jacquinot} to reach shorter $t_c$, which should improve the
utility of our model for proton NMR experiments. 

Related effects can occur for a wider variety of $\mathcal H_{int}$ and $\mathcal H_{P_\phi}$ than we have treated here,
provided that $[\mathcal H_{P_\phi},\mathcal H_{int}]\neq 0$. Shaped pulses, soft pulses, and strongly-modulated pulses
have proven to be important elements of the NMR toolbox.  Exploiting the internal structure of hard $\pi$ pulses provides
us with yet another technique to control the coherent evolution of quantum systems.

We thank K. W. Zilm, E. K. Paulson, and R. Tycko for discussions, and thank M. H. Devoret, S. M. Girvin, A. Pines, and C. P. Slichter for comments on the manuscript.  
This work was supported by the NSF under grants DMR-0207539, DMR-0325580, and DMR-0653377.

%%%%%%%%%%%%%%%%%%%%%%%%%%%
%%%%%%  E N D    B O D Y    T E X T %%%%%%%%
%%%%%%%%%%%%%%%%%%%%%%%%%%%


\begin{references}
\bibitem{Slichter}C. P. Slichter, {\em Principles of Magnetic Resonance} (Springer, New York, 1990), 3rd ed.
\bibitem{Mehring}M. Mehring, {\em Principles of High Resolution NMR in Solids} (Springer-Verlag, Berlin, 1983), 2nd ed.
\bibitem{Ernst}R. R. Ernst, G. Bodenhausen, and A. Wokaun, {\em Principles of Nuclear Magnetic Resonance in One and Two
Dimensions} (Clarendon, Oxford, 1987).
\bibitem{DalePRL} Dale Li, A. E. Dementyev, Yanqun Dong, R. G. Ramos and S. E. Barrett, Phys. Rev. Lett. {\bf 98},
190401 (2007).
\bibitem{DalePRB} Dale Li, Yanqun Dong, R. G. Ramos, J. D. Murray, K. MacLean, A. E. Dementyev,  and S. E. Barrett,
Phys. Rev. B, accepted  (available at http://arxiv.org/abs/0704.3620)
\bibitem{Purcell} H.Y. Carr, and E. M. Purcell, Phys. Rev. {\bf 94}, 630 (1954)
\bibitem{Gill} S. Meiboom, and D. Gill, Rev. Sci. Instrum. {\bf 29}, 688 (1958)
\bibitem{WaughPR} U. Haeberlen, and J. S. Waugh, Phys. Rev. {\bf 175}, 453 (1968)
\bibitem{Hahn}E. L. Hahn,  Phys. Rev. {\bf 80}, 580 (1950).
\bibitem{WaughPRL} W. K. Rhim, A. Pines,  and J. S. Waugh, Phys. Rev. Lett. {\bf 25}, 218 (1970).
\bibitem{WaughPRB} W. K. Rhim,  A. Pines, and J. S. Waugh, Phys. Rev. B {\bf 3}, 684 (1971).
\bibitem{McDowell} K.Takegoshi, and C. A. McDowell,  Chem. Phys. Lett. {\bf 116}, 100 (1985).
\bibitem{WaughJMR} A. Pines, and J. S. Waugh, J. Mag. Res. {\bf 8}, 354 (1972).
\bibitem{WaughJCP} U. Haeberlen, J. D. Ellett, and J. S. Waugh,  J. Chem. Phys. {\bf 55}, 53 (1971).
\bibitem{RonaPRB} R. G. Ramos, Yanqun Dong, Dale Li, and S. E. Barrett (in preparation).
\bibitem{Solomon} I. Solomon, Phys. Rev. Lett. {\bf 2}, 301 (1959).
\bibitem{Matsui} S. Matsui, Chem. Phys. Lett. {\bf 179}, 187 (1991).
\bibitem{Garroway} J. B. Miller, D. G. Cory,  and A. N. Garroway, Phil. Trans. R. Soc. Lond. A {\bf 333}, 413 (1990).
\bibitem{Blumich} D. E. Demco,  and B. Blumich, Concepts Magn. Reson. {\bf 12}, 269 (2000).
\bibitem{Mansfield} P. Glover and P. Mansfield, Rep. Prog. Phys. {\bf 65}, 1489 (2002).
\bibitem{Ackerman} Y. Wu, D. A. Chesler, M. J. Glimcher, L. Garrido, J. Wang, H. J. Jiang, and J. L. Ackerman, Proc.
Natl. Acad. Sci. USA {\bf 96}, 1574 (1999).
\bibitem{Glimcher} Y. Wu,  J. L. Ackerman, H. M. Kim, C. Rey, A. Barroug, and M. J. Glimcher, J. Bone Miner. Res. {\bf
17}, 472 (2002).
\bibitem{Lauterbur} T. L. Peck, R. L. Magin, and P. C. Lauterbur,  J. Magn. Reson. B {\bf 108}, 114 (1995).
\bibitem{Kentgens} K. Yamauchi, J. W. G. Janssen, and A. P. M. Kentgens,  J. Magn. Reson. {\bf 167}, 87 (2004).
\bibitem{Jacquinot} D. Sakellariou, G. Le Goff,  and J. F. Jacquinot, Nature {\bf 447}, 694 (2007).

\end{references}

\end{document}